\documentclass[aip,pof,amsmath,amssymb,reprint,
superscriptaddress,
]{revtex4-1}

\usepackage{graphicx}
\usepackage{dcolumn}
\usepackage{bm}
\usepackage{CJK}
\usepackage{epstopdf}
\usepackage{color}

\begin{document}

\preprint{AIP/POF}

\title{Horizontal Structures of Velocity and Temperature Boundary Layers in 2D Numerical Turbulent Rayleigh-B\'{e}nard Convection}

\author{Quan ZHOU}
\email{qzhou@shu.edu.cn (Q.Z.); kxia@phy.cuhk.edu.hk (K.Q.X.). Authors to whom correspondence should be addressed.}
\affiliation{Shanghai Institute of Applied Mathematics and Mechanics, Shanghai University, Shanghai 200072, China}
\affiliation{Shanghai Key Laboratory of Mechanics in Energy and Environment Engineering and Modern Mechanics Division in E-Institutes of Shanghai Universities, Shanghai University, Shanghai 200072, China}

\author{Kazuyasu SUGIYAMA}
\affiliation{Physics of Fluids Group, Faculty of Science and Technology, J. M. Burgers Centre for Fluid Dynamics, and Impact-Institute, University of Twente, 7500 AE Enschede, The Netherlands}
\affiliation{Department of Mechanical Engineering, School of Engineering, The University of Tokyo, Tokyo, Japan}

\author{Richard J. A. M. STEVENS}
\affiliation{Physics of Fluids Group, Faculty of Science and Technology, J. M. Burgers Centre for Fluid Dynamics, and Impact-Institute, University of Twente, 7500 AE Enschede, The Netherlands}

\author{Siegfried GROSSMANN}
\affiliation{Fachbereich Physik, Philipps-Universit\"{a}t Marburg, D-35032 Marburg, Germany}

\author{Detlef LOHSE}
\affiliation{Physics of Fluids Group, Faculty of Science and Technology, J. M. Burgers Centre for Fluid Dynamics, and Impact-Institute, University of Twente, 7500 AE Enschede, The Netherlands}

\author{Ke-Qing XIA}
\email{qzhou@shu.edu.cn (Q.Z.); kxia@phy.cuhk.edu.hk (K.Q.X.). Authors to whom correspondence should be addressed.}
\affiliation{Department of Physics, The Chinese University of Hong Kong, Shatin, Hong Kong, China}

\date{\today}

\begin{abstract}
We investigate the structures of the near-plate velocity and temperature profiles at different horizontal positions along the conducting bottom (and top) plate of a Rayleigh-B\'{e}nard convection cell, using two-dimensional (2D) numerical data obtained at the Rayleigh number Ra$=10^8$ and the Prandtl number Pr$=4.4$ of an Oberbeck-Boussinesq flow with constant material parameters. The results show that most of the time, and for both velocity and temperature, the instantaneous profiles scaled by the dynamical frame method [Q. Zhou and K.-Q. Xia, {\it Phys. Rev. Lett.} {\bf 104}, 104301 (2010)] agree well with the classical Prandtl-Blasius laminar boundary layer (BL) profiles. Therefore, when averaging in the dynamical reference frames, which fluctuate with the respective instantaneous kinematic and thermal BL thicknesses, the obtained mean velocity and temperature profiles are also of Prandtl-Blasius type for nearly all horizontal positions. We further show that in certain situations the traditional definitions based on the time-averaged profiles can lead to unphysical BL thicknesses, while the dynamical method also in such cases can provide a well-defined BL thickness for both the kinematic and the thermal BLs.

\end{abstract}


\maketitle

\section{Introduction}

In a series of recent studies \cite{zx2010prl, zss2010jfm} we have experimentally and numerically analyzed the structures of the kinematic and thermal boundary layers (BLs) in the vicinity of the horizontal top and bottom plates, where the fluid layer is heated from below and cooled from above in turbulent Rayleigh-B\'{e}nard (RB) convection, in the central region of the RB cell \cite{agl2009rmp, lx2010arfm}. The dynamics and the global features of the thermal convection system are strongly influenced, sometimes even dominated by the properties of the BL flow. Nearly all theories of the heat transport in turbulent RB convection, from the early marginal stability theory \cite{malkus1954} to the Shraiman $\&$ Siggia (SS) model \cite{ss1990pra, siggia1994arfm} and to the Grossmann $\&$ Lohse (GL) theory \cite{gl2000jfm, gl2001prl, gl2002pre, gl2004pof}, are essentially BL theories. Therefore, it is a key issue of turbulent RB convection, how the near-plate velocity and temperature profiles look like.

Specifically, the GL theory has achieved great success in predicting the global quantities, such as the Rayleigh number \cite{funfschilling2005jfm} and Prandtl number \cite{xia2002prl} dependence of the heat flux, i.e., the Nusselt number, and amplitude of the large scale circulation (LSC), i. e., the Reynolds number, of the turbulent RB system \cite{agl2009rmp}. Recently, the GL theory was successfully extended to the very large Rayleigh number regime (the so called ultimate range), in order to predict the experimentally observed multiple scaling of the heat transfer \cite{gl2011pof} and to the rotating case to predict the heat transfer enhancement \cite{scl2010pof}. As the GL theory is based on the assumption that the BL thickness scales inversely proportional to the square root of the Reynolds number according to Prandtl's 1904 theory, the validity of  Prandtl-Blasius BL flow needs to be tested also locally. Note that comparison of the mean bulk temperature calculated using the Prandtl-Blasius theory with that measured in both liquid and gaseous non-Oberbeck-Boussinesq RB convection shows very good agreement \cite{agl2006jfm, agl2007prl, agl2008pre}. In addition, the kinematic BL thickness evaluated by solving the laminar Prandtl-Blasius BL equations was found to agree well with that obtained in the direct numerical simulation (DNS) \cite{ssg2010njp}.

\subsection{BLs along the cell's central axis}

Previous works about the kinematic and thermal BLs \cite{zx2010prl, zss2010jfm} mainly focused on the cell's central vertical axis. Although the  flow in the bulk is turbulent, the BLs are considered to behave still laminar at least scaling wise because of the small shear Reynolds number in the BLs (see Fig. \ref{fig:fig7}(b) of this paper for the values of the shear Reynolds numbers in the present case). Indeed, in a time-averaged sense, it was found experimentally that the kinematic BL thicknesses $\lambda_v$ near the sidewalls of a cubic cell \cite{qiu1998prea} and near the bottom plate of a rectangular cell \cite{sun2008jfm} obey the Prandtl scaling for a laminar flat plate BL, i.e. $\lambda_v\sim Re^{-1/2}$, where $Re$ is the Reynolds number of the LSC in the RB system. Furthermore, certain wall quantities, such as the wall shear stress, the friction velocity, and the viscous sublayer thickness, were also found to follow the Prandtl scaling \cite{sun2008jfm}. However, direct comparisons of experimental velocity \cite{thess2007prl} and numerical temperature \cite{thess2009jfm} profiles with the respective classical Prandtl-Blasius profiles show significant deviations, especially for the distances from the plate around the BL thickness. It was argued that such deviations should be attributed to the intermittent emissions of thermal plumes from the BLs and the corresponding temporal dynamics of the BLs \cite{sugiyama2009jfm, zx2010prl}. This led to the study of the BL structures in dynamical reference frames, which fluctuate with the instantaneous BL thicknesses, rather than in the laboratory frame \cite{zx2010prl}. When resampling the velocity and temperature fields' data in such dynamical frames, both the mean velocity and the temperature profiles were found to agree well with the respective theoretical Prandtl-Blasius laminar BL profiles over a wider parameter range of both Ra and Pr \cite{zx2010prl, zss2010jfm}. Moreover, when the instantaneous velocity and temperature profiles are rescaled by their respective instantaneous BL thicknesses, it was found that the Prandtl-Blasius profiles not only hold in a time-averaged sense, but are most of the time also valid in an instantaneous sense \cite{zss2010jfm}. A dynamical BL rescaling method has thus been established, which extends the time-independent Prandtl-Blasius BL theory to the time-dependent case, in the sense that it holds locally at every instant in the frame that fluctuates with the local instantaneous BL thickness. All this was, as mentioned, shown for the center range of the RB cell.

\subsection{The spatial structures of the BLs}

\begin{figure}
\begin{center}
\resizebox{0.95\columnwidth}{!}{%
  \includegraphics{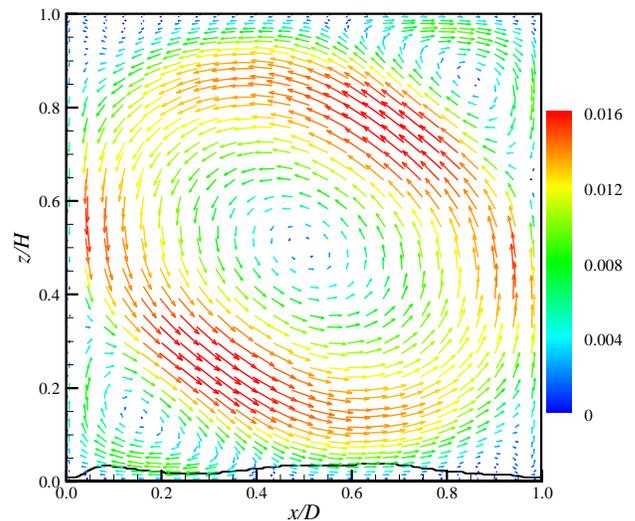}
}
\caption{(color online). The time-averaged vector map of the whole velocity field $\vec{v} = (u,w)$ (see Sec. \ref{sec:rev} for the details). For clarity, a coarse-grained vector map of size $26\times50$ meshpoints is shown. The magnitude of the velocity $v = \sqrt{u^2+w^2}$ is coded in both color and the length of the arrows in units of m/s. The time average is taken over a period of 80 min corresponding to 30 000 velocity maps and to 500 large eddy turnovers (LET), with $T_{LET} = (2H + 2D) / u_{LSC} = 9.6$ s, $u_{LSC} = 0.017 $ms$^{-1}$, and 60 velocity maps per LET. If one follows the stream trace passing through the maximal velocity $u_{LSC}$ and uses the numerically measured local velocities along the approximately elliptically shaped circumference of the stream strace, the turnover time is only $7.2$ s, corresponding to 667 LET in $80$ min. -- The solid curve marks the kinematic BL thickness near the bottom plate. The Reynolds number of the LSC is $Re_{LSC}=u_{LSC}H/\nu=1036$, where $u_{LSC}$ is the maximal velocity magnitude of the LSC, and the Reynolds number of the lower left corner roll is $Re_{cr}=u_{cr}\ell_{cr}/\nu=134$, where $u_{cr} = 0.011$ ms$^{-1}$ is the maximal velocity magnitude of the lower left corner roll and $\ell_{cr}\simeq0.2H$ is the typical length scale for the corner roll. } \label{fig:fig1}
\end{center}
\end{figure}

\begin{figure}
\begin{center}
\resizebox{0.95\columnwidth}{!}{%
  \includegraphics{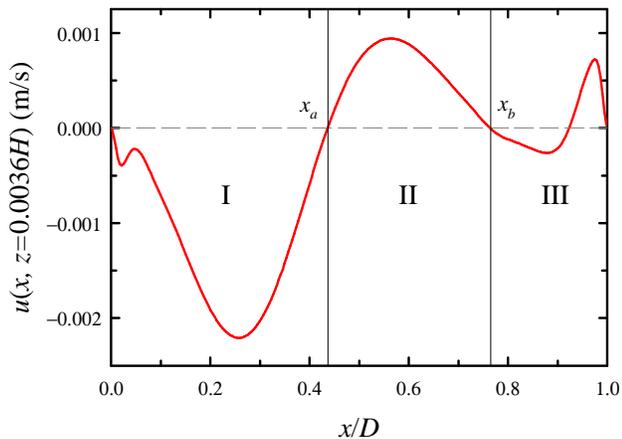}
}
\caption{(color online). The time-averaged horizontal velocity profile $u(x)$ as a function of $x/D$ obtained near the bottom plate ($z/H=0.0036$). The profile can be divided into three regions: the left corner roll (region I), the LSC (region II), and the right corner rolls (region III). The vertical solid lines mark the boundaries between the three regions at $x_a$ and $x_b$. Note that at $Ra=1\times10^8$ both the kinematic and thermal BL thicknesses are larger than $0.009H$ for all horizontal positions, see the solid curve in Fig. \ref{fig:fig7} (a). Thus the horizontal velocity profile in this figure is well within both BLs.
}
\label{fig:fig2}
\end{center}
\end{figure}

As a closed system, turbulent thermal convection in an RB cell develops rather complicated flow structures, partly due to the interactions between the flow and the solid walls, cf. \cite{sugiyama2009jfm}. The kinematic and thermal BLs along the cell's central vertical axis thus cannot reveal all BL properties, especially not for the BLs in the regions near the cell's corners. In Fig. \ref{fig:fig1}, we show an example of the time-averaged vector map of the whole velocity field obtained from a two-dimensional (2D) simulation with the Rayleigh number Ra$=10^8$ and the Prandtl number Pr$=4.4$ (for the details of the simulations we refer to Sec. \ref{sec:setup}). As usual the Rayleigh number is defined as $Ra \equiv \alpha g H^3 \Delta/\nu\kappa$ and the Prandtl number as Pr$\equiv\nu/\kappa$. Here $\nu$, $\kappa$, $\alpha$, and $g$ are the kinematic viscosity, thermal diffusivity, isobaric thermal expansion coefficient, and gravitational acceleration; $H$ denotes the height of the container and $\Delta$ the temperature difference between the hotter bottom and the cooler top temperature. It is seen clearly that the overall flow pattern is an counter- clockwise rotatory motion. While in a three-dimensional (3D) cylindrical cell \cite{sun2005pre} the mean flow was found to be elliptically shaped, the large-scale circulation (LSC) in the present case looks a bit more stadium-like shaped with its long and short axes pointing approximately to the cell's two diagonals. There are several smaller secondary rolls at the four corners of the cell: two larger clockwise rolls at the two opposite corners adjacent to the short axis of the LSC ellipse and much smaller vortices at the two opposite corners adjacent to the long axis of the LSC ellipse. Thus the flow near the horizontal plates can be divided into the two corner-roll regions and the central region dominated by the LSC. To see this more clearly, we plot in Fig. \ref{fig:fig2} the horizontal profile of the time-averaged horizontal velocity $u(x)$ near the bottom plate ($z=0.0036H$). One can distinguish three different ranges of $x$ that differ from each other by different values of $u(x)$. Region I is the left-corner-roll region where the flow is dominated by the clockwise corner roll and $u(x)$ is negative, region II is the central region where the flow is dominated by the counter-clockwise LSC and $u(x)$ is positive. The transition point between region I and II is identified as $x_a/D=0.44$. In region III the flow is somewhat complicated and both negative and positive $u(x)$ are observed, indicating several small vortices in the right corner. The transition point between region II and III is identified as $x_b/D=0.77$. -- On the upper plate the velocity profile is correspondingly, but from right to left.

Such complicated flow structures near the horizontal plates highlight the need to study the horizontal dependence of the local BL profiles, both of the velocity and the temperature. Compared to the large amount of studies on the near-plate velocity and temperature profiles along the cell's central vertical axis, however, studies on the spatial dependence of these profiles off the center line are very limited both experimentally or in simulations.

The spatial structure of the thermal BL in water in the range $10^{8}<\text{Ra}<10^{10}$ has systematically been studied first by Lui and Xia in a cylindrical cell \cite{lui1998pre} and then by Wang and Xia in a cubic cell \cite{wang2003epjb}. Both experiments have shown that the thermal BL thickness above the bottom plate, $\lambda_{th}$, depends on the horizontal position $x$ along the plate, and the scaling exponent of $\lambda_{th}$ with Ra varies between $-0.35$ and $-0.28$. However, this position-dependence is expected to decrease with increasing Ra, i.e., $\lambda_{th}$ tends to eventually become uniform along the plate at very large Ra. This behavior can be shown to result from the shape evolution of the LSC. Namely, its shape evolves from a tilted and nearly elliptical shape at low Ra to a more squarish shape at high Ra \cite{niemela2003epl, sun2005pre}. The squarish-shaped LSC at high Ra will make the mean flow near the horizontal plates to be more parallel to the plates and hence leads to more  uniformity of the BLs.

The spatial structures of the kinematic BL in water was first studied experimentally by Qiu and Xia \cite{qiu1998preb} in a cubic cell. It was found that the magnitudes of the LSC, the shear rate, and the kinematic BL thickness all change significantly across the horizontal plates both parallel  and perpendicular to the LSC. Direct comparison between the observed temperature BL profiles and the Prandtl-Blasius thermal profiles at different horizontal positions were performed by Sugiyama \emph{et al.} \cite{sugiyama2009jfm} in 2D and by Stevens \emph{et al.} \cite{stevens2010jfm} in 3D numerical simulations. It was found that due to the rising (falling) plumes near the sidewalls the deviations of the numerically calculated BL profiles from the Prandtl-Blasius profiles increase from the center of the horizontal plates towards the sidewalls.

\subsection{The objective of the present work}

In this paper we want to extend previous works \cite{zx2010prl, zss2010jfm} dealing with the plate's center to the whole bottom (top) plate with the help of 2D DNS. Our results will show that the idea of the dynamical BL thickness rescaling method works well for almost all horizontal positions, i.e., the mean BL profiles obtained at nearly all horizontal positions can be brought into coincidence with the Prandtl-Blasius laminar BL profiles, if they are re-sampled in the time-dependent frames of the local BL thicknesses, for both velocity and temperature.

\section{Definitions, numerical parameters, and data analysis}
\label{sec:setup}

\subsection{Numerical methods}

The mathematical model, the numerical scheme, and the code validation have been described elsewhere \cite{sugiyama2009jfm}. Thus we give only their main features here. The computational domain consists of a 2D square cell of horizontal length $D=4.078$ cm and hight $H=4.078$ cm, the aspect ratio is thus $\Gamma\equiv D/H=1$. The flow is calculated by numerical integration of the 2D time-dependent imcompressible Oberbeck-Boussinesq equations with a fourth-order finite-difference scheme. No-slip velocity boundary conditions are applied to all four solid walls. As temperature boundary conditions the two sidewalls are chosen to be adiabatic (no flux), while at the colder top and the warmer bottom plates the temperatures are fixed. The mean temperature is chosen as $T_m = 40^{\circ}$C and water as the working fluid. Then the kinematic viscosity, thermal diffusivity, and isobaric thermal expansion coefficient are $\nu=6.6945\times10^{-7}$ m$^2$s$^{-1}$, $\kappa=1.5223\times10^{-7}$ m$^2$s$^{-1}$, and $\alpha=3.8343\times10^{-4}$ K$^{-1}$. The resulting Prandtl number is Pr $\equiv\nu/\kappa=4.4$. During the computation, the temperature difference across the fluid layer was fixed at $\Delta=40$ K. The corresponding Rayleigh number then is Ra $\equiv \alpha g H^3 \Delta /\nu\kappa=10^8$.

For these values of the control parameters the Nusselt number $Nu$ (in its usual definition as a $z$-independent area average; here in DNS after $z$-averaging in addition) is $Nu = 25.62$, which corresponds to the $x$-independent overall thermal BL thickness $\lambda_{th} / H = 1 / 2Nu = 0.0195$. If $Nu$ is calculated with the slopes of the area (here the $x$) averages $\partial_z \langle \theta \rangle_x$ at the bottom or top plates, the respective values are $25.65$ and $25.69$, being within $0.3$\% with the $z$-average.

We denote the computational domain as the $(x,z)$-plane. Then at any time $t$ the horizontal and vertical velocity components, $u(x,z,t)$ and $w(x,z,t)$, and the temperature $\theta(x,z,t)$ are obtained. For the temperature field a non-dimensional temperature $\Theta(x,z,t)$ is introduced, namely,
\begin{equation}
\Theta(x,z,t)=\frac{\theta_{bot}-\theta(x,z,t)}{\Delta/2}.
\end{equation}
Here $\theta_{bot}$ is the fixed hot temperature at the bottom plate. Based on this definition, the fixed dimensionless temperatures of the top and bottom plates are $\Theta(x,H,t)=2$ and $\Theta(x,0,t)=0$, respectively, and the mean bulk temperature is $\Theta=1$.

\subsection{Reversals of the LSC}
\label{sec:rev}

\begin{figure}
\begin{center}
\resizebox{0.95\columnwidth}{!}{%
  \includegraphics{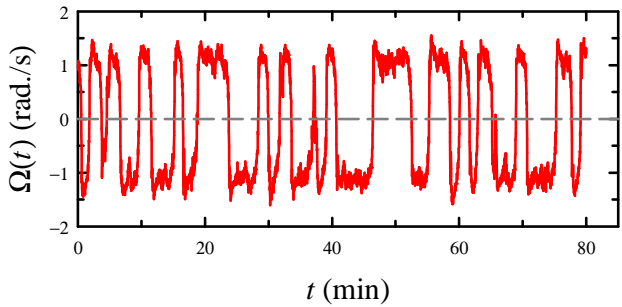}
}
\caption{(color online). Time trace of $\Omega(t)$.} \label{fig:fig3}
\end{center}
\end{figure}

Previous studies have shown that for the numerical parameters used in the present work, the mean flow of the 2D RB convection would experience spontaneous flow reversals, due to the competitions between the corner rolls and the LSC \cite{sns2010prl}. This can be characterized by the global angular velocity of the whole flow field, defined as
\begin{equation}
\label{eq:vor}
\Omega(t) \equiv \left< \frac{w(x,z,t)}{x-0.5D}-\frac{u(x,z,t)}{z-0.5H}\right>_s,
\end{equation}
where $\langle\cdots\rangle_s$ denotes the spatial, i.e., ($x,z$) average. Based on the definition of $\Omega(t)$, $\Omega(t)>0$ indicates the counter-clockwise rotation of the mean flow while the mean flow rotates clockwise when $\Omega(t)<0$. The events of flow reversals can therefore be identified through a sign change of $\Omega(t)$. Figure \ref{fig:fig3} shows a time series of $\Omega(t)$. One can identify a total of 32 reversals in a period of 80 minutes. Such a flow with so many reversals in the relevant averaging time yields a nearly zero time mean velocity field, because the counter-clockwise and clockwise velocity contributions cancel under time-averaging. To overcome the influence of the reversals, before analyzing the BL data we transform the 2D velocity and temperature fields, from $u(x,z,t)$, $w(x,z,t)$, and $\Theta(x,z,t)$ to $u'(x,z,t)$, $w'(x,z,t)$, and $\Theta'(x,z,t)$, as follows:
\begin{enumerate}
  \item at any time $t$, if $\Omega(t)\geq0$ (counter-clockwise), the 2D velocity and temperature fields are taken unchanged, i.e. $u'(x,z,t)=u(x,z,t)$, $w'(x,z,t)=w(x,z,t)$, and $\Theta'(x,z,t)=\Theta(x,z,t)$;
  \item at any time $t$, if $\Omega(t)<0$ (clockwise), the 2D velocity and temperature fields are reflected with respect to the axis $x/D=0.5$, i.e., $u'(x,z,t)=-u(D-x,z,t)$, $w'(x,z,t)=w(D-x,z,t)$, and $\Theta'(x,z,t)=\Theta(D-x,z,t)$.
\end{enumerate}
After such transformation, the LSC (if it exists; during the reversals the LSC breaks down \cite{sns2010prl}) would rotate counter-clockwise for all times $t$. A resulting time-averaged velocity field is illustrated in Fig. \ref{fig:fig1}. For convenience, in the remainder of the paper we also use the notations of $u(x,z,t)$, $w(x,z,t)$, and $\Theta(x,z,t)$ for the velocity and temperature after the described transformation. Since the governing equations are strictly Oberbeck-Boussinesq (temperature independent fluid parameters), one expects top-bottom symmetry. Though for theoretical reasons this must be valid, we also have checked that numerically for the velocity field explicitly. We found complete agreement in all details of the $x$-dependent structures. All respective profiles collapse under top-bottom and $x \rightarrow D-x$ mapping. We thus need to discuss only the velocity and the temperature profiles near the bottom plate (without labeling that, since it also holds for the top plate, respectively).

\subsection{The kinematic and thermal BL thicknesses obtained from the time-averaged profiles}
\label{sec:tablt}

\begin{figure}
\begin{center}
\resizebox{0.95\columnwidth}{!}{%
  \includegraphics{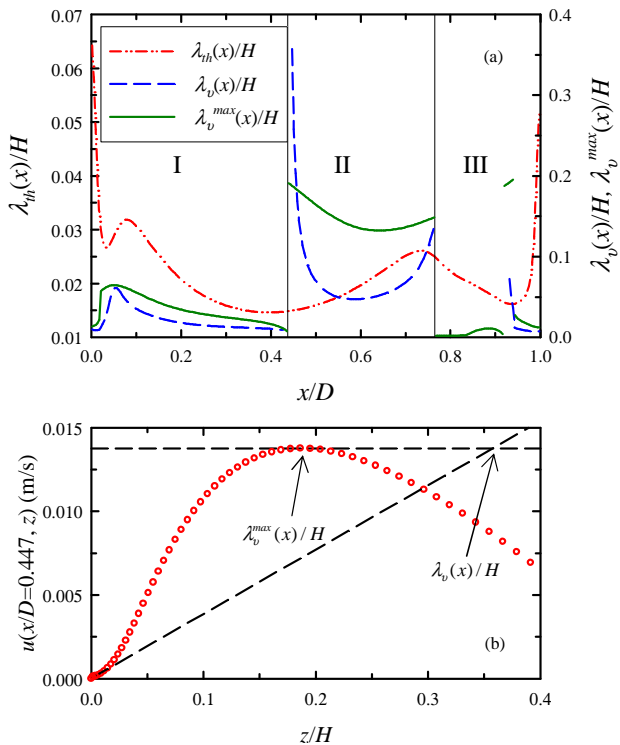}
}
\caption{(color online). (a) Normalized kinematic and thermal BL thicknesses in $z$-direction near the bottom plate, $\lambda_v(x)/H$, $\lambda_v^{max}(x)/H$, and $\lambda_{th}(x)/H$, as functions of the horizontal position $x$, obtained from the time-averaged velocity and temperature fields $u(x,z)$ and $\Theta(x,z)$. The vertical solid lines mark the boundaries $x_a$ and $x_b$ between the three regions. Note the different scales used for the kinematic and thermal BL thicknesses. (b) The time-averaged velocity $z$-profile $u(x,z)$ for fixed $x/D=0.447$ (near the boundary $x_a$). The tilted dashed line is a linear fit to the near-plate part of the $u$-velocity's $z$-profile and the horizontal dashed line marks the maximum horizontal velocity.} \label{fig:fig4}
\end{center}
\end{figure}

With the measured time-averaged velocity and temperature fields, the kinematic and thermal BL thicknesses $\lambda_v(x)$ and $\lambda_{th}(x)$ can be defined, respectively, via the $z$-slopes of the time-averaged velocity and temperature $z$-profiles for each horizontal position $x$,
\begin{equation}
u(x,z)=\langle u(x,z,t)\rangle \mbox{\ and\ } \Theta(x,z)=\langle \Theta(x,z,t)\rangle,
\end{equation}
cf. \cite{agl2009rmp, tbl1993pre}, where $\langle\cdots\rangle$ denotes the time average. Usually the width $\lambda_v^{99\%}$ is considered instead. Since in RB geometry the velocity does not level asymptotically but decreases again with $z$, we here consider $\lambda_v^{max}$ or $\lambda_{th}^{max}$, defined as the width at which the profile reaches its (first) maximum. The max- and the 99\%-widths differ only minutely.

Figure \ref{fig:fig4}(a) shows $\lambda_v(x)/H$ and $\lambda_{th}(x)/H$ as functions of $x/D$. The thermal BL is much thicker at the two corners because of the rising plumes near the sidewalls. The horizontal distribution of $\lambda_{th}(x)$ is asymmetric and there exists a minimum value of $\lambda_{th}(x)$ at $x/D\simeq0.4$. These features of the thermal BL thickness as a function of $x$ are similar to those observed in a cubic cell \cite{wang2003epjb} and to those in a numerical study of a square cell \cite{werne1993pre}. In contrast, in a cylindrical cell a symmetric ``$\vee$" shape of the $\lambda_{th}(x)$ profile was found \cite{lui1998pre}. This difference was attributed to the effects of the sharp corners in the cubic or square cells on the flow \cite{wang2003epjb}, suggesting an impact of the cell geometry on the characteristics of thermal BL structures in turbulent RB convection.

The $x$-dependence of $\lambda_v(x)$ is much more complicated. Here one finds large discontinuities and even gaps in the measured $\lambda_v(x)$ profile around the two boundaries $x_a$ and $x_b$ of the three regions. These features are caused by the strong competition between the corner rolls and the LSC at these positions and apparently are unphysical. The fluctuating instantaneous border between the corner rolls and the LSC makes the velocity change its sign [see Fig. \ref{fig:fig8}(c)] and slope near the plate very quickly, so that a kinematic BL thickness cannot be well defined. Figure \ref{fig:fig4}(b) shows an example of the time-averaged velocity $z$-profile obtained at fixed $x/D=0.447$ (near the boundary $x_a$). At this $x$-position, the strong competition between the corner and center rolls makes the near-plate velocities sometimes positive (when the flow is dominated by the LSC) and sometimes negative (when the flow is dominated by the left corner roll). The positive and negative velocities cancel each other when calculating the time-averaged velocities, yielding a rather small velocity slope near the plate (see the tilted dashed line in the figure) and thus a very large value of $\lambda_v(x)$, which is obviously unphysical. Therefore, the traditional definition of the kinematic BL thickness from the slope of the time-averaged $z$-profile cannot properly handle the situation as shown in Fig. \ref{fig:fig4}. One may define, instead, the kinematic BL thickness as the distance $\lambda_v^{max}(x)$ to the near-plate extremal horizontal velocity. However, still the sign change of the  velocity along the horizontal direction [see Fig. \ref{fig:fig2}] produces unphysical discontinuities and large jumps in the $\lambda_v^{max}(x)/H$ profile [see Fig. \ref{fig:fig4}(a)]. As we shall see below, these unphysical results for the time-averaged BL thickness can be eliminated if we use the instantaneous BL thickness based on the instantaneous velocity profiles.

We have also checked the profiles and thicknesses of the rms-temperature and -velocity profiles (cf. \cite{sugiyama2009jfm}). $\lambda_{th}^{rms}(x)$ and the slope thickness $\lambda_{th}(x)$ show the same features as functions of $x$, the rms-thickness being about 20\% below the slope thickness. The velocity field's BL thickness $\lambda_v^{rms}(x)$ is much closer to $\lambda_v^{max}(x)$, within 5\%; its $x$-dependence reflects the same structures, including the discontinuities. Only the boundaries $x_{a,b}$ are somewhat shifted. We thus do not consider the rms-fields further.

\subsection{The instantaneous kinematic and thermal BL thicknesses}

\begin{figure}
\begin{center}
\resizebox{0.95\columnwidth}{!}{%
  \includegraphics{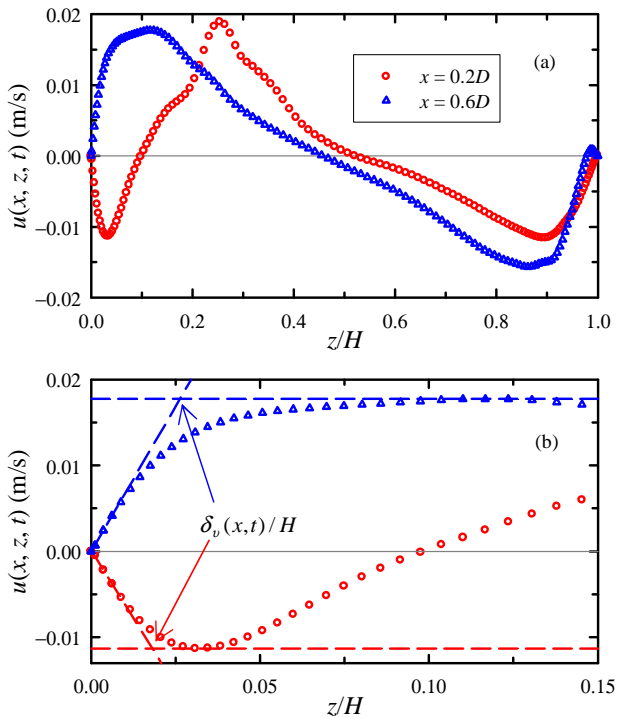}
}
\caption{(color online). (a) Examples of the instantaneous horizontal velocity's $u(x,z,t)$ $z$-profiles obtained at $x=0.2D$ (red circles) and $x=0.6D$ (blue triangles). (b) An enlarged part of the velocities' $z$-profiles near the bottom plate. The tilted dashed lines are linear fits to the linear part of the velocity profiles near the bottom plate and the horizontal dashed lines mark the instantaneous minimum (for $x=0.2D$) or maximum (for $x=0.6D$) horizontal velocities near the bottom plate. The distances of the crossing points from the plate define the instantaneous local kinematic BL thicknesses $\delta_v(x,t)$, either of the LSC or of the corner roll.} \label{fig:fig5}
\end{center}
\end{figure}

\begin{figure}
\begin{center}
\resizebox{0.95\columnwidth}{!}{%
  \includegraphics{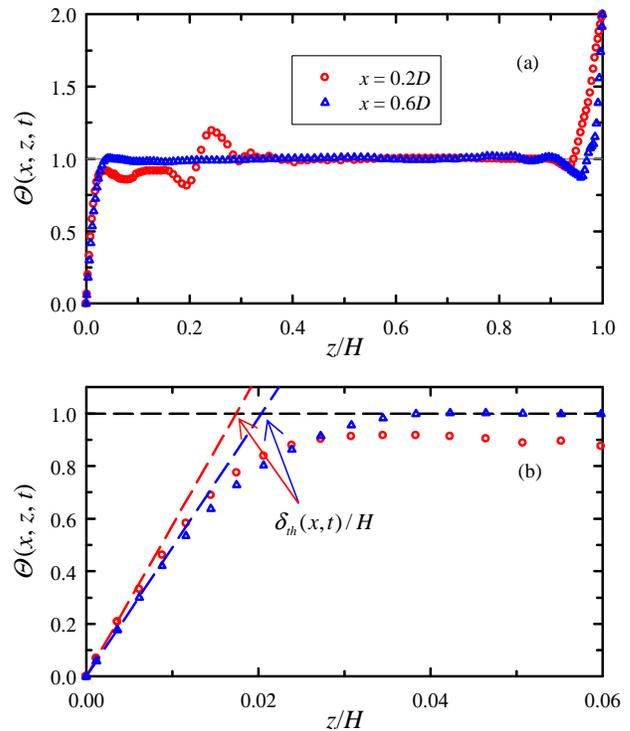}
}
\caption{(color online). (a) Examples of the normalized instantaneous local temperature $z$-profiles $\Theta(x,z,t)$ obtained at $x=0.2D$ (red circles) and $x=0.6D$ (blue triangles). (b) An enlarged part of the temperature profiles near the bottom plate. The tilted dashed lines are linear fits to the linear parts of the temperature $z$-profiles near the bottom plate and the horizontal dashed line mark the bulk temperature $\Theta=1$. The distances of the crossing points from the plate define the instantaneous local thermal BL thicknesses $\delta_{th}(x,t)$. The latter one here is smaller for $x = 0.2 D$ than for $x = 0.6 D$.} \label{fig:fig6}
\end{center}
\end{figure}

Figure \ref{fig:fig5}(a) shows examples of the instantaneous horizontal velocity's $u(x,z,t)$ vertical profiles versus $z$ for fixed horizontal position $x$ and time $t$, in particular at $x/D=0.2$ and $x/D=0.6$. As shown in Figs. \ref{fig:fig1} and \ref{fig:fig2}, in the time-averaged sense position $x/D=0.6$ belongs to region II where the flow is dominated by the LSC. The properties of $u(x=0.6D,z,t)$ are similar to those reported in our previous works \cite{zx2010prl, zss2010jfm}. Namely, $u(x=0.6D,z,t)$ increases very quickly from 0 to the instantaneous maximum velocity within a very thin layer above the bottom plate and then decreases slowly in the bulk region of the closed convection cell. In contrast, the $z$-profile $u(x=0.2D,z,t)$ in region I shows quite different features. Here the flow is dominated by the left corner roll. It is seen that $u(x,z,t)$ as a function of $z$ first drops very quickly from 0 to the instantaneous minimum velocity near the bottom plate, i.e., the velocity very near to the plate is negative, the flow is towards the corner. After reaching its near-plate (negative) minimum value, $u(x=0.2D,z,t)$ rises very quickly from negative values through 0 to the instantaneous maximum velocity and then decreases slowly in the bulk region. The kinematic BL at this position is produced and stabilized by the viscous shear of the corner roll, rather than by the LSC. The total thickness of the corner or secondary roll at this time is of the order $0.4 H$. Figure \ref{fig:fig5}(b) shows the enlarged near-plate parts of the two velocity profiles. One sees that both profiles are linear near the bottom plate and thus the instantaneous local kinematic BL thickness can be defined as the distance from the plate at which the extrapolation of the linear part of $u(x,z,t)$ versus $z$ crosses the horizontal line passing through the instantaneous near-plate extremum horizontal velocity (e.g. minimum for $x=0.2D$ and maximum for $x=0.6D$). We denote these instantaneous local BL thicknesses by the symbol $\delta_v(x,t)$ to distinguish them from the thicknesses $\lambda_v$ (or $\lambda_{th}$) of the time-averaged $z$-profiles. The dashed lines in Fig. \ref{fig:fig5}(b) illustrate how to determine $\delta_v(x,t)$ as the crossing point distances for the two $x$-positions.

Figure \ref{fig:fig6}(a) shows examples of the $z$-profiles of the normalized instantaneous local temperature $\Theta(x,z,t)$, obtained at fixed $x=0.2D$ and $x=0.6D$ and chosen time $t$. Unlike the case of $u(x,z,t)$ in Fig. \ref{fig:fig5}, although the two positions are dominated by  different flow directions, both $\Theta(x,z,t)$ here exhibit similar features as functions of $z$. $\Theta(x,z,t)$ increases very quickly within a very thin layer above the bottom plate and stays nearly constant at the mean bulk temperature $\Theta=1$, valid in the bulk region. Figure \ref{fig:fig6}(b) shows an enlarged near-plate part of the two temperature profiles. Linear pieces can be seen in both profiles near the bottom plate. The instantaneous local thermal BL thickness (again denoted by the symbol $\delta$ instead of $\lambda$) $\delta_{th}(x,t)$ can then be defined as that  distance from the plate, where the extrapolation of the linear part of $\Theta(x,z,t)$ crosses the horizontal line passing through the mean bulk temperature, even if the measured temperature stays below that, as for $x = 0.2 D$. The dashed lines in Fig. \ref{fig:fig6}(b) illustrate how to determine $\delta_{th}(x,t)$ as the crossing point distances.

\begin{figure}
\begin{center}
\resizebox{0.95\columnwidth}{!}{%
  \includegraphics{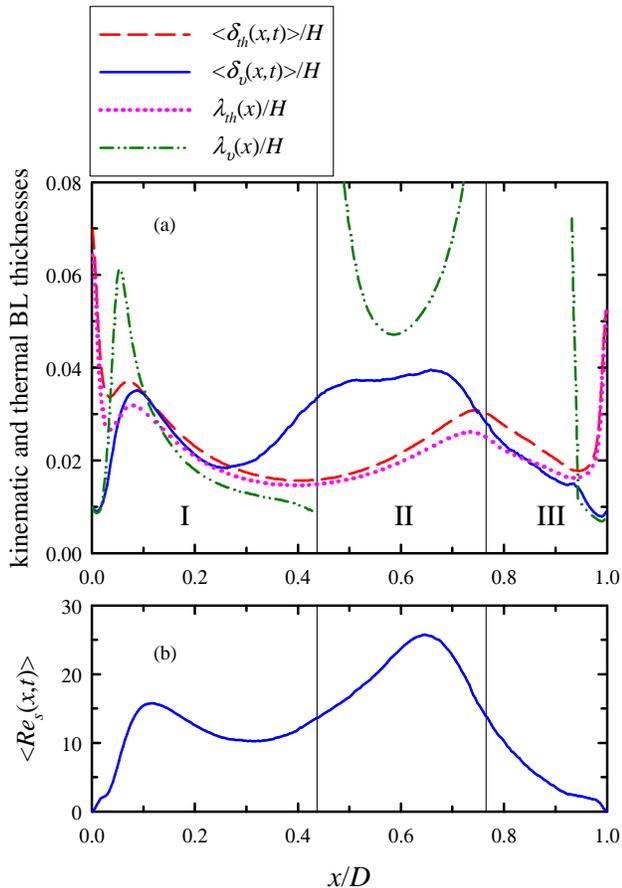}
}
\caption{(color online). (a) The horizontal ($x$-)dependence of the time-averaged local instantaneous BL widths $\langle\delta_v(x,t)\rangle/H$ and $\langle\delta_{th}(x,t)\rangle /H$. The dashed curves mark the horizontal dependence of $\lambda_{th}(x)/H$ and $\lambda_v(x)/H$ for comparison. (b) The time-averaged values of the instantaneous shear Reynolds number $\langle Re_s(x,t)\rangle$ as a function of $x/D$. The vertical solid lines mark the boundaries $x_a$ and $x_b$ between the three regions I, II, III.} \label{fig:fig7}
\end{center}
\end{figure}

As discussed in Sec. \ref{sec:tablt}, the kinematic local BL thicknesses, $\lambda_v(x)$ and $\lambda_v^{max}(x)$, based on the time-averaged velocity profiles have some unphysical features. Specifically, Fig. \ref{fig:fig4}(b) illustrated the limits of the traditional definition of the BL thickness. We now show that the use of the instantaneous BL can avoid some problems.  We consider the time-averaged mean values of the instantaneous local BL thicknesses $\langle\delta_v(x,t)\rangle$ and $\langle\delta_{th}(x,t)\rangle$  as typical measures of the local kinematic and thermal BL thicknesses, respectively. Figure \ref{fig:fig7} shows $\langle\delta_v(x,t)\rangle/H$ and $\langle\delta_{th}(x,t)\rangle/H$ as functions of $x/D$. For comparison,  $\lambda_{th}(x)/H$ and $\lambda_v(x)/H$ obtained from the time-averaged temperature and velocity profiles are also plotted as dashed curves in the figure. For temperature, one sees that the horizontal dependences of $\langle\delta_{th}(x,t)\rangle/H$ and $\lambda_{th}(x)/H$ share the same trend, with $\langle\delta_{th}(x,t)\rangle/H$ only a little bit larger than $\lambda_{th}(x)/H$. For velocity, the situation is quite different. Note that the time-averaged local instantaneous BL width $\langle\delta_v(x,t)\rangle/H$ varies smoothly between 0.008 and 0.04 along the whole horizontal plate, especially around the boundaries $x_a$ and $x_b$ between the different regions. No gaps exist and no extremely large thicknesses. This is because at any instant the instantaneous local kinematic BL thickness $\delta_v(x,t)$ can be well defined regardless whether the local flow is dominated by the LSC or by the corner rolls at that particular instant. Therefore, $\langle\delta_v(x,t)\rangle/H$ can be used to characterize the typical length scale of the kinematic BLs in situations where $\lambda_v(x)/H$ and $\lambda_v^{max}(x)/H$, based on the time-averaged velocity profiles, are no longer capable of producing physically meaningful results.

Let us briefly compare with the conventional BL thickness. The area- (here the $x$-)averaged thermal BL thickness is $\langle \delta_{th}(x) \rangle_x = 0.024 H$; this is slightly but definitely different from the global thickness $\lambda_{th} = H / 2Nu = 0.0195 H$ given in Sec. II.A. The $x$-averaged rms-profile thickness $\langle \lambda_{th}^{rms}(x) \rangle_x = 0.0176 H$ turns out to be less. -- For the kinematic BL, in contrast, due to the corner rolls and the corresponding back flows, an $x$-averaged $v$-profile is not meaningful. But one might $x$-average the local thicknesses to obtain $\langle \delta_v (x) \rangle_x = 0.0267 H$. Compare this average with the $x$-dependent thicknesses in Fig. 4(a) and find that under averaging the smaller corner roll BLs reduce the center roll BL thickness considerably. -- Note that the slope thickness of the center roll is of order $\lambda_v \approx a / \sqrt{Re_{LSC}}$ with $a \approx 1.6$. This differs from the value $a = 0.5$ valid in the Prandtl law for the global $Re$ number, cf. \cite{gl2000jfm, gl2001prl, gl2002pre, gl2004pof}.

The instantaneous local BL thicknesses also yield the so-called local shear Reynolds number, based on the local kinematic BL thickness as the characteristic length scale. We define it as
\begin{equation}
Re_s(x,t)=\frac{u(x,\delta_v,t)\delta_v(x,t)}{\nu},
\end{equation}
where $u(x,\delta_v,t)$ is the instantaneous local velocity at $z=\delta_v(x,t)$. Figure \ref{fig:fig7}(b) shows the horizontal variation of the time-averaged values of the instantaneous local shear Reynolds numbers, $\langle Re_s(x,t)\rangle$. The $x$-profile of this time averaged local shear Reynolds numbers shows two peaks. One peak is within the lower left corner roll and the other is within the LSC. These local shear Reynolds numbers  $\langle Re_s(x,t)\rangle$ are for all horizontal positions $x$ much smaller than the critical value $Re_s=420$ for the instability of the boundary layer, which had been proposed in the literature cf. \cite{landau1987}. This suggests a still laminar though temporally fluctuating BL for the present control parameters Ra and Pr. Note that in a previous experimental study \cite{sun2008jfm} it was found from the extrapolation of the $Re_s$ versus Ra scaling that for $Pr=4.3$ a turbulent BL is expected to occur at Ra$\simeq2\times10^{13}$ (the similar conclusion was also made in an experimental study of local heat flux measurements \cite{shang2008prl}).

\subsection{The dynamical BL rescaling method}

With the measured $\delta_v(x,t)$ and $\delta_{th}(x,t)$, the local dynamical kinematic and thermal BL frames at different horizontal positions $x$ along the bottom or top plates can now be constructed. We define the time-dependent relative vertical distances $z^*_v(x,t)$ and $z^*_{th}(x,t)$ from the plate with respect to $\delta_v(x,t)$ and $\delta_{th}(x,t)$, respectively, as
\begin{equation}
z^*_v(x,t)\equiv\frac{z}{\delta_v(x,t)} \mbox{\ \ and\ \ } z^*_{th}(x,t)\equiv\frac{z}{\delta_{th}(x,t)}.
\end{equation}
The mean local velocity and temperature profiles, $u^*(x,z_v^*)$ and $\Theta(x,z^*_{th})$, in the respective dynamical BL frames at any horizontal position $x$ are then defined as
\begin{equation}
u^*(x,z^*_v)\equiv\langle |u(x,z=z^*_v\delta_v(x,t),t)|\rangle
\end{equation}
and
\begin{equation}
\Theta^*(x,z^*_{th})\equiv\langle \Theta(x,z=z^*_{th}\delta_{th}(x,t),t)\rangle,
\end{equation}
i.e. time-averaging over all values of $|u(x,z,t)|$ and $\Theta(x,z,t)$ that were measured at different discrete times $t$, but at the same rescaled positions $z^*_v$ and $z^*_{th}$, respectively. Here we use the absolute values $|u(x,z,t)|$ when calculating the mean velocity profiles, because $u(x,z,t)$ has different signs at different horizontal positions $x$ and at different time $t$, especially for the positions around $x_a$ and $x_b$, and $u(x,z,t)$ with different signs would cancel each other partially.

In order to characterize the shapes of the (time-averaged or instantaneous) local velocity and temperature $z$-profiles and to study their agreement or deviations from the respective Prandtl-Blasius profiles, both quantitatively, we compute the local shape factors $H_i(x)$ of the profiles, cf. \cite{schlichting2004}
\begin{equation}
H_i(x)\equiv\delta_i^d(x)/\delta_i^m(x), \mbox{\ \ with\ \ } i=v, th,
\end{equation}
where $\delta_i^d(x)$ and $\delta_i^m(x)$ are the local displacement and momentum thicknesses of the profiles, respectively, defined as,
\begin{equation}
\delta^d_i(x)\equiv\int_0^{\infty} \left[1-\frac{Y(x,z)}{[Y(x,z)]_{max}}\right] dz
\end{equation}
and
\begin{equation}
\delta^m_i(x)\equiv\int_0^{\infty}\ \left[1-\frac{Y(x,z)}{[Y(x,z)]_{max}}\right] \left[\frac{Y(x,z)}{[Y(x,z)]_{max}}\right] dz.
\end{equation}
Here, $Y(x,z)=|u(x,z)|$ is horizontal local velocity's $z$-profile if $i=v$, and $Y(x,z)=\Theta(x,z)$ is the corresponding local temperature's $z$-profile if $i=th$. As suggested in our previous studies \cite{zx2010prl, zss2010jfm}, all $z$-integrations are evaluated only over the range from $z=0$ to the position of the first maximum of the $z$-profiles instead of $\infty$. The shape factor of a profile describes how fast the profile approaches its asymptotic value. The larger the shape factor is, the faster the profile runs to its asymptotic level. The shape factor of the Prandtl-Blasius velocity profile is $H_v^{PB}=2.59$ independent of Pr and that of the Prandtl-Blasius temperature profile is $H_{th}^{PB}=2.61$ for the present Pr$=4.4$. The deviations of the numerical $z$-profiles from the respective Prandtl-Blasius profiles can then be measured by
\begin{equation}
\delta H_i=H_i-H_i^{PB}.
\end{equation}

\section{Velocity profiles near the plate}

\begin{figure}
\begin{center}
\resizebox{0.8\columnwidth}{!}{%
  \includegraphics{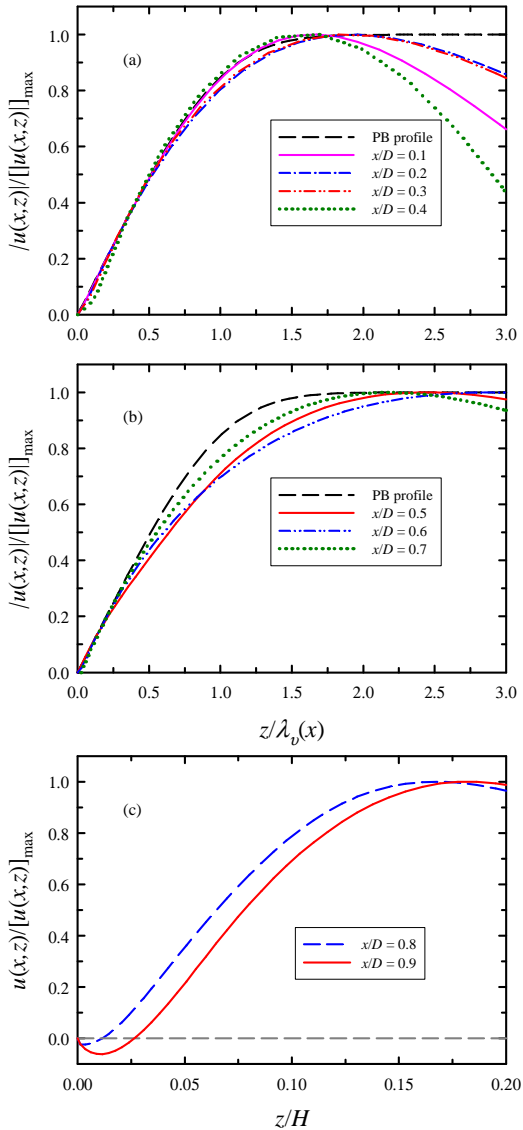}
}
\caption{(color online). (a, b) The absolute values of the time-averaged local horizontal velocity $z$-profiles, $|u(x,z)|$, as functions of the normalized distance $z/\lambda_v(x)$ obtained at (a) $x/D=0.1$, 0.2, 0.3, and 0.4 (region I) and (b) $x/D=0.5$, 0.6, and 0.7 (region II). Here, $|u(x,z)|$ is normalized by its respective maximum value near the bottom plate, $[|u(x,z)|]_{max}$. The dashed lines indicate the Prandtl-Blasius velocity profile for comparison. (c) The time-averaged local horizontal velocity $z$-profiles $u(x,z)$, normalized by the respective maximum horizontal velocity near the bottom plate $[u(x,z)]_{max}$, as functions of $z/H$ obtained at $x/D=0.8$ and 0.9 (region III).} \label{fig:fig8}
\end{center}
\end{figure}

\begin{figure}
\begin{center}
\resizebox{0.8\columnwidth}{!}{%
  \includegraphics{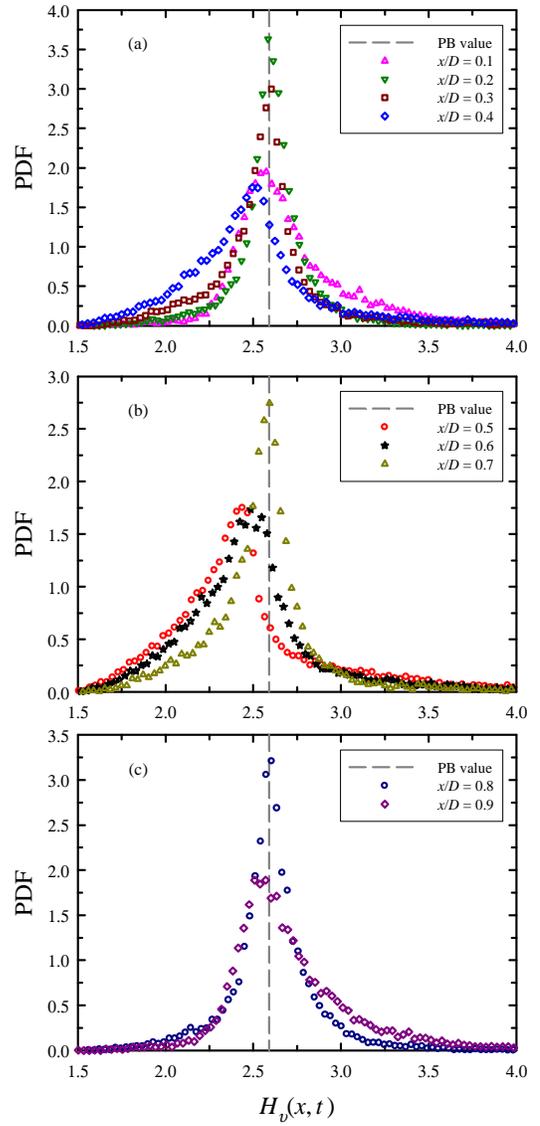}
}
\caption{(color online). PDFs of the shape factors $H_v(x,t)$ of the rescaled instantaneous velocity profiles, see figure \ref{fig:fig10}, obtained at (a) $x/D=0.1$, 0.2, 0.3, and 0.4 (region I), (b) $x/D=0.5$, 0.6, and 0.7 (region II), and (c) $x/D=0.8$ and 0.9 (region III). The dashed lines mark the shape factor of the Prandtl-Blasius velocity profile for comparison.} \label{fig:fig9}
\end{center}
\end{figure}

\begin{figure*}
\begin{center}
\resizebox{1.9\columnwidth}{!}{%
  \includegraphics{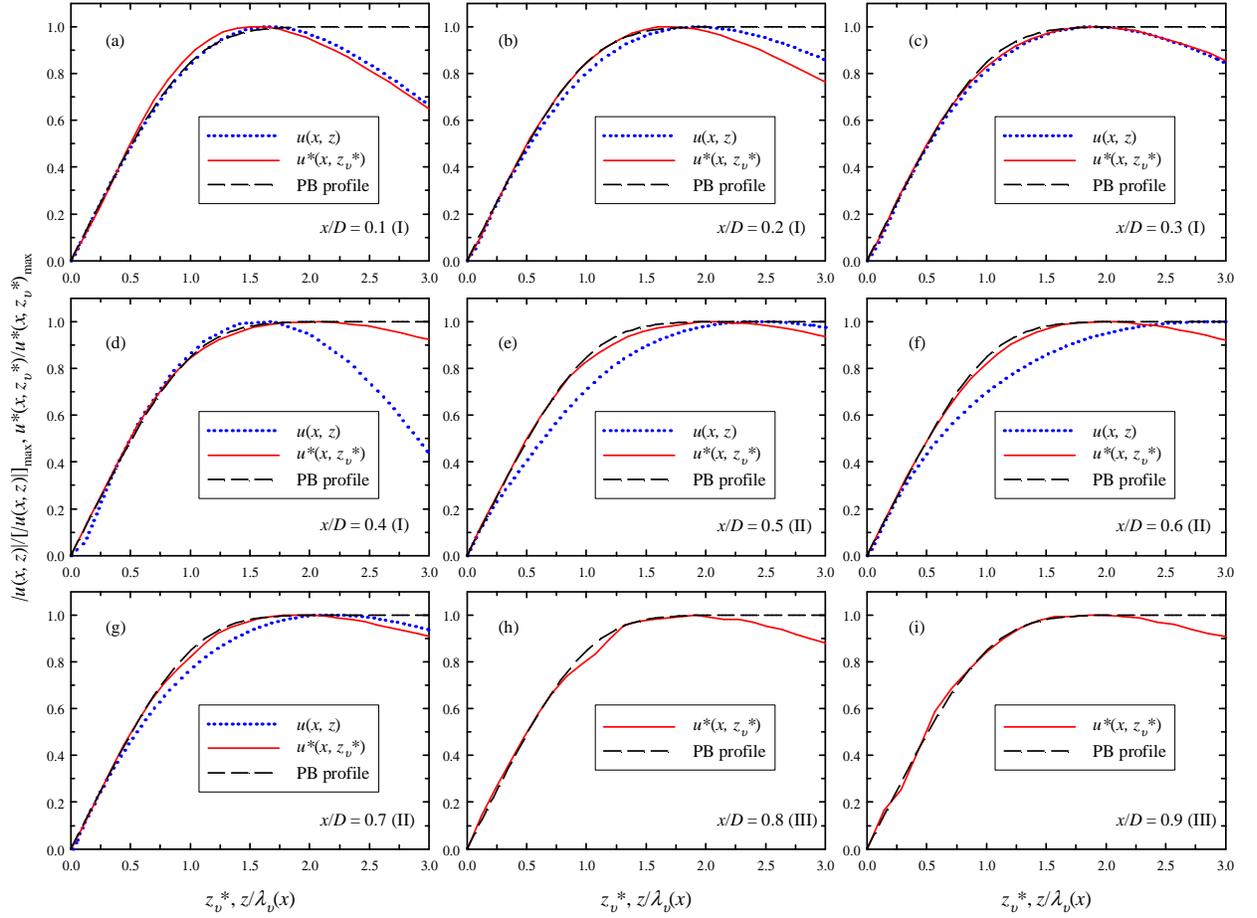}
}
\caption{(color online). Comparison between velocity profiles obtained at $x/D=0.1$, 0.2, ..., 0.9 near the bottom plate: dynamical $u^*(z_v^*)$ (red solid lines), laboratory $u(z)$ (blue solid lines), and the Prandtl-Blasius laminar velocity profile (black dashed lines). Note that the Prandtl-Blasius profile per construction stays constant once it has reached its asymptotic value, while the RB flow profiles decrease towards the bulk. Thus agreement can only be expected in the very BL region. As the plots show, in the BL range the dynamically rescaled instantaneous local profiles are very well consistent with the Prandtl-Blasius shapes.} \label{fig:fig10}
\end{center}
\end{figure*}

\begin{figure}
\begin{center}
\resizebox{0.95\columnwidth}{!}{%
  \includegraphics{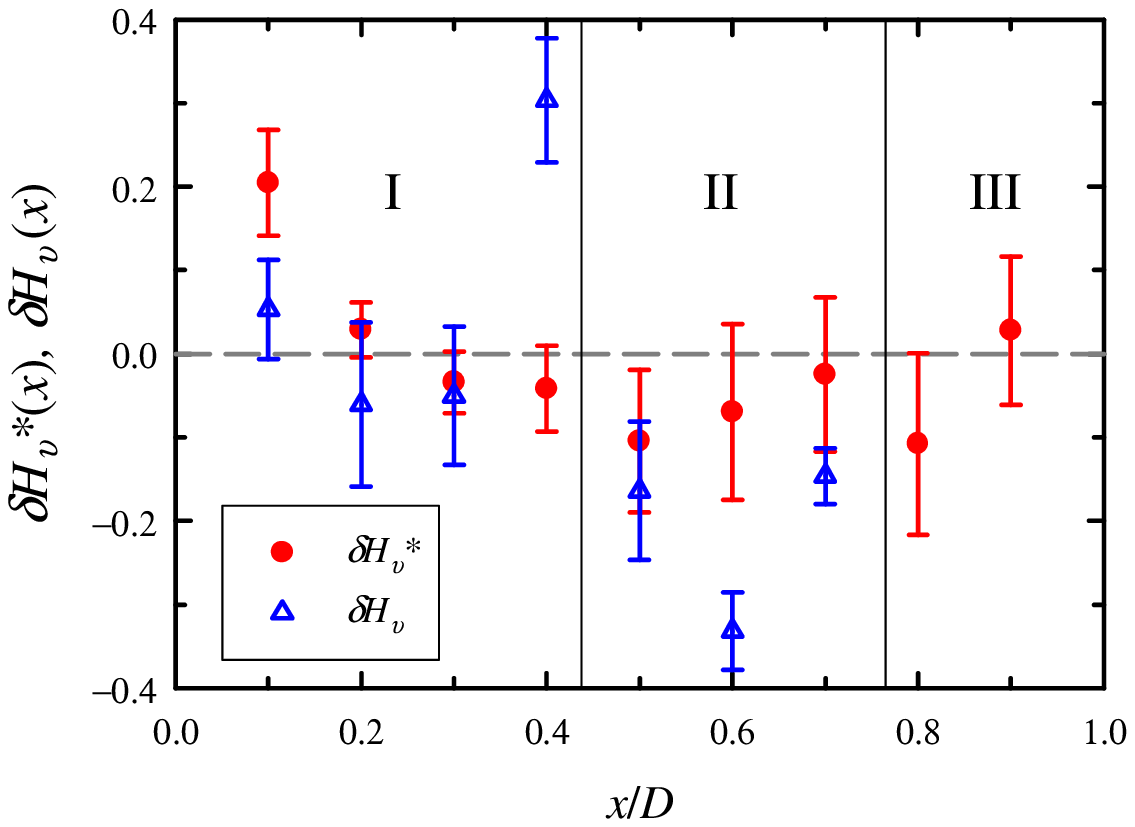}
}
\caption{(color online). The horizontal ($x$-)dependence of the deviations of the local velocity profile shape factors from the Prandtl-Blasius value for the laboratory frame, $\delta H_v(x)$ (open triangles) and the time averaged dynamical frame, $\delta H_v^*(x)$ (solid circles). The vertical solid lines mark the boundaries between the three regions I, II, III.} \label{fig:fig11}
\end{center}
\end{figure}

To reveal the horizontal ($x$-)dependence of the BL structures using the dynamical rescaling method, we focus on nine different horizontal positions at $x/D=i\times0.1$ with $i=1$, 2, ..., and 9. As shown in Figs. \ref{fig:fig1} and \ref{fig:fig2}, in the time-averaged sense the positions $x/D=0.1$, 0.2, 0.3, and 0.4 belong to region I, where the flow is dominated by the left corner roll. The positions $x/D=0.5$, 0.6, and 0.7 belong to region II, where the flow is dominated by the LSC. The positions $x/D=0.8$ and 0.9 belong to region III, where the flow is dominated by the small right corner rolls. To reduce the data scatter, before applying the dynamical rescaling method we coarse grain the local horizontal velocity and temperature profiles $u(x,z,t)$ and $\Theta(x,z,t)$ obtained at each position $i$ and at each discrete time $t$ by averaging them along the $x$-direction (horizontally) over the range $i\times0.1-0.01<x/D<i\times0.1+0.01$.

Figures \ref{fig:fig8}(a) and (b) show the magnitudes of the $z$-profiles of the time-averaged local horizontal velocity $|u(x,z)|$, obtained respectively in regions I and II. Here, the velocities are normalized by their respective near-plate maximum values $[|u(x,z)|]_{max}$ and the vertical distance $z$ is  normalized by the local kinematic BL thickness $\lambda_v(x)$. For comparison, we also plot the Prandtl-Blasius velocity $z$-profile (the dashed lines), the initial slope of which is matched to those of the measured profiles. It is seen that the time-averaged velocity $z$-profiles measured at positions within region II deviate significantly from the Prandtl-Blasius profile (Fig. \ref{fig:fig8}(b)). On the other hand the $z$-profiles in region I, where the flow is dominated by the left corner roll, match the Prandtl-Blasius one much better (Fig. \ref{fig:fig8}(a)). This observation was somewhat to our surprise, since for the temperature profiles in both 2D \cite{sugiyama2009jfm} and 3D \cite{stevens2010jfm} simulations increasing deviations from the Prandtl-Blasius profile were found when away from the cell center, due to the rising plumes close to the sidewall. Our present result may be understood from Fig. \ref{fig:fig1}, in which one sees that the mean flow (in the corner roll) in region I is essentially parallel to the plate, which is just the case treated by the Prandtl-Blasius BL theory, i.e., a horizontal flow over a flat plate. In contrast, the flow enters region II from above not parallel to the plate and only becoming more horizontal afterwards.

Figure \ref{fig:fig8}(c) shows the time-averaged local velocity profile $u(x,z)$ including its sign, measured at $x/D=0.8$ and 0.9, i.e., in region III. The velocities $u(x,z)$ are still normalized by their respective maximum values, while $z$ is normalized here by the cell's height $H$ instead of the local BL widths. Due to the very small right corner rolls in region III, the two profiles first drop a little from 0 and then increase to their maximum values. We find that there are respectively 1 and 4 data points between 0 and the near-plate minimum velocities of the two profiles. These numbers of data points are so small that it is meaningless to define BL thicknesses for these profiles. In the present study, the kinematic BL thickness was calculated only, when the number of data points between 0 and the near-plate extremal velocities was larger than 5.

Having analyzed the profiles of the time-averaged local velocities with the tools of the time-averaged fields and have noted considerable deviations from the Prandtl-Blasius behavior, we now consider its comparison with the $z$-profiles of the instantaneous local horizontal velocity. At first we evaluate the local instantaneous shape factors $H_v(x,t)$ to characterize the respective $z$-profiles of the horizontal velocity near the plate, see Fig. \ref{fig:fig9}. Then the dynamically rescaled $z$-profiles themselves will be considered, see Fig. \ref{fig:fig10}. The corresponding analysis for the thermal profiles will be presented in Sec. IV.

Figure \ref{fig:fig9}(a) shows the probability density functions (PDFs) of the local shape factors $H_v(x,t)$ of the $z$-profiles of the instantaneous local velocity, obtained at $x/D=0.1$, 0.2, 0.3, and 0.4 within region I. The dashed vertical line in the figure denotes the Prandtl-Blasius value for comparison. It is seen that the distributions are exactly peaked at the Prandtl-Blasius value, except at the position $x/D=0.4$, where the peak is slightly off. This illustrates that most of the time the instantaneous local velocity profiles in region I are indeed of Prandtl-Blasius type. Note that the position $x/D=0.4$ is close to the boundary $x_a$ between the regions I and II and hence the peak's slight deviation from the Prandtl-Blasius value at this position is likely to be caused by the competition between the corner roll and the LSC.

Figure \ref{fig:fig9}(b) shows the PDFs of $H_v(x,t)$ obtained at $x/D=0.5$, 0.6, and 0.7 within region II. One sees that the peak positions of the distributions move closer to $H_v^{PB}$ when proceeding from position $x/D=0.5$, where the LSC is still slighly tilted downwards flowing, to the more plate parallel flow position of the LSC at $x/D=0.7$. Specifically, at position $x/D=0.7$, the PDF is exactly peaked at the Prandtl-Blasius value. This illustrates that the $z$-profiles of the instantaneous local velocity are becoming more Prandtl-Blasius along the evolution of the LSC. Referring to Fig. \ref{fig:fig1} this may be understood as follows. The Prandtl-Blasius BL starts to develop in this region from the boundary between the corner roll and the LSC near $x/D=0.4$, and as one moves downstream the LSC becomes stronger and steadier so that it produces a more Prandtl-Blasius-like laminar layer. -- The PDFs of $H_v(x,t)$ measured at $x/D=0.8$ and 0.9 within region III are plotted in Fig. \ref{fig:fig9}(c). Again, the two distributions are peaked close to $H_v^{PB}$, indicating that the instantaneous local velocity profiles obtained at these positions are of Prandtl-Blasius type for most of the time.

We now consider the $z$-profiles of the instantaneous local horizontal velocity in direct comparison with the profiles of the times averaged fields seen from the laboratory system. Figure \ref{fig:fig10} shows this direct comparison among the velocity profiles obtained at nine different horizontal positions $x_i$: the dynamical frame based local instantaneous horizontal velocity $u^*(x,z_v^*)$, the laboratory frame based time-averaged velocity profile $u(x,z)$, and the Prandtl-Blasius kinematic BL profile. Overall, obviously the $u^*(x,z_v^*)$ profiles obtained in the dynamical frames match the Prandtl-Blasius profile well. This can be understood from our results in Fig. \ref{fig:fig9} that most of the time the instantaneous velocity profiles are of Prandtl-Blasius type and hence averaging all the rescaled profiles in the dynamical BL frames would naturally yield a profile of Prandtl-Blasius type.

Figures \ref{fig:fig10}(a) to (d) show the profiles obtained in region I. Both $|u(x,z)|$ and $u^*(x,z_v^*)$ approximately match the Prandtl-Blasius profile for the range $z_v^*\lesssim2$, i.e., in the proper BL range. This suggests that the plume flow and temporal dynamics of the BLs do not play a key role in this region, which may be attributed to the strong mixing between the corner roll and the LSC. Note that after reaching the maximum values, both $|u(x,z)|$ and $u^*(x,z_v^*)$ decrease towards the bulk of the closed convection cell, while the Prandtl-Blasius profile keeps unchanged because it describes the situation of an asymptotically constant, non-zero flow velocity.

Figures \ref{fig:fig10}(e) to (g) show the velocity profiles measured in region II. One observes that the time-averaged profiles, $|u(x,z)|$, obtained in the laboratory frame is much lower than the Prandtl-Blasius profile in the region around the kinematic BL thickness. As discussed in our previous papers \cite{zx2010prl, zss2010jfm}, a simple average of velocities at a fixed height $z$ in the laboratory frame will sample a mixed dynamics, one pertaining to the BL range and the other one pertaining to the bulk, owing to the fluctuations of the BL thickness, and thus will distort the shapes of the profiles from that of Prandtl-Blasius. In contrast, the $u^*(x,z_v^*)$ profiles measured in the instantaneous local dynamical frame agree well with the Prandtl-Blasius velocity profile, suggesting that the dynamical BL rescaling method can effectively disentangle the mixed dynamics of the BLs and the bulk, as all profiles are expressed in the intrinsic BL-length scale.

Figures \ref{fig:fig10}(h) and (i) show the velocity profiles obtained in region III. Because of the small negative part of the profiles near the plate (see Fig. \ref{fig:fig8}(c)), we better plot $u^*(x,z_v^*)$ instead of $|u(x,z)|$ in the figures for comparison. One observes that   $u^*(x,z_v^*)$ averaged in the dynamical BL frames also here agree well with the Prandtl-Blasius $z$-profile, despite the existence of very complicated corner rolls in this region.

The quantitative deviations of the velocity profiles shown in Fig. \ref{fig:fig10} from the Prandtl-Blasius profile in terms of the shape factors are plotted in Fig. \ref{fig:fig11}. When time-averaging in the laboratory frame, the velocity shape-factor deviations $\delta H_v(x)$ are closed to 0 near the cell sidewall ($x/D=0.1$, 0.2, and 0.3), but far away from 0 at the other positions. In contrast, the shape-factor deviations $\delta H_v^*(x)$ of the velocity profiles in the dynamical time-dependent local frames obviously are all much closer to zero, except that at $x/D=0.1$, which is a bit of over-corrected. These quantitative results further indicate that our dynamical BL rescaling method works well for nearly all horizontal positions. And they confirm the interesting fact that the Prandtl-Blasius laminar BL theory in RB flow is well justified despite the significant time dependence due to a vivid plume dynamics for Rayleigh numbers Ra below the ultimate state.

\begin{figure}
\begin{center}
\resizebox{0.8\columnwidth}{!}{%
  \includegraphics{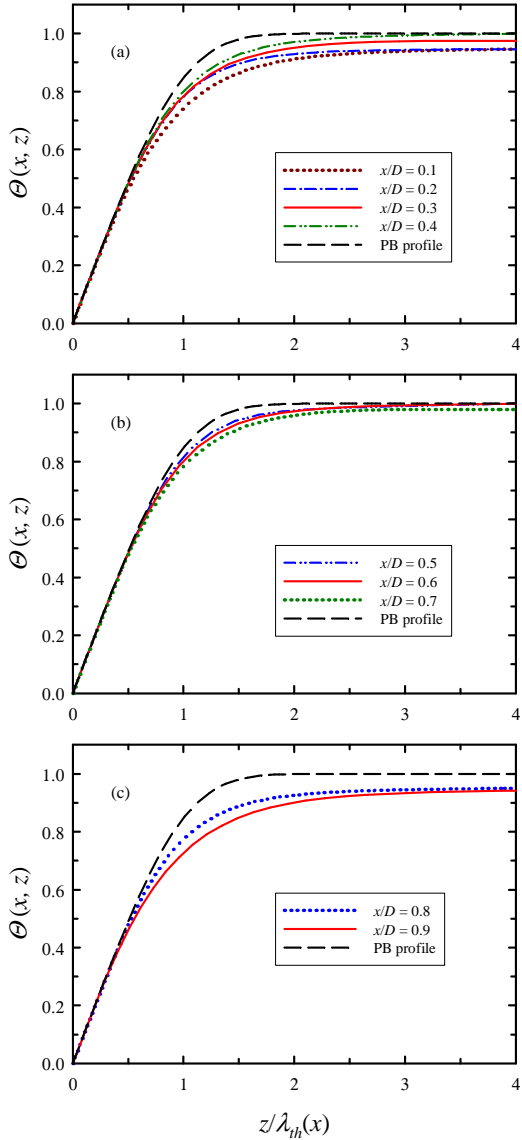}
}
\caption{(color online). (a) The $z$-profiles of the time-averaged temperature field $\Theta(x,z)$ as functions of the normalized distance $z/\lambda_{th}(x)$ using the lab-frame local thicknesses, obtained in (a) region I at  $x/D=0.1$, 0.2, 0.3, and 0.4; in (b) region II at  $x/D=0.5$, 0.6, and 0.7; in (c) region III at  $x/D=0.8$ and 0.9. The dashed lines indicate the Prandtl-Blasius thermal profile for comparison. } \label{fig:fig12}
\end{center}
\end{figure}

\begin{figure}
\begin{center}
\resizebox{0.95\columnwidth}{!}{%
  \includegraphics{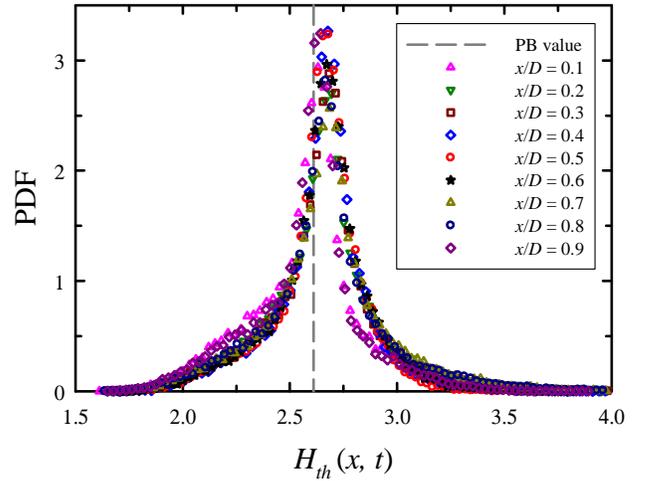}
}
\caption{(color online). PDFs of the shape factors of the rescaled instantaneous local temperature profiles (profiles saturate at first maximum, see text for explanation) obtained at $x/D=0.1$, 0.2, ..., 0.9. The dashed line marks the shape factor of the Prandtl-Blasius thermal profile for comparison.} \label{fig:fig13}
\end{center}
\end{figure}

\begin{figure*}
\begin{center}
\resizebox{1.9\columnwidth}{!}{%
  \includegraphics{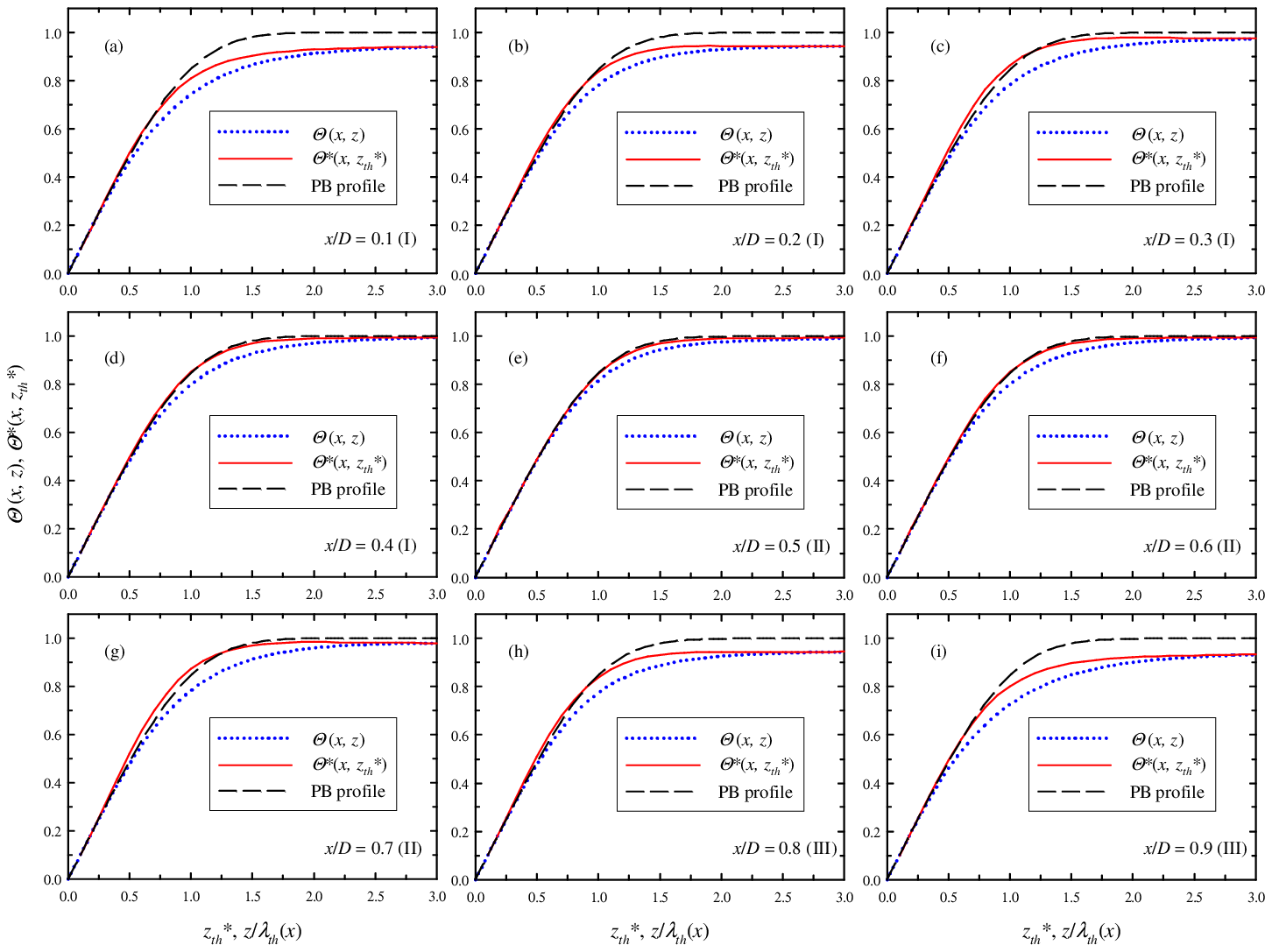}
}
\caption{(color online). Comparison between temperature $z$-profiles obtained at $x/D=0.1$, 0.2, ..., 0.9 near the bottom plate: dynamical $\Theta^*(x,z_{th}^*)$ (red solid lines), laboratory $\Theta(x,z)$ (blue solid lines), and the Prandtl-Blasius laminar thermal profile (black dashed lines), see also figure \ref{fig:fig15}. }\label{fig:fig14}
\end{center}
\end{figure*}

\begin{figure*}
\begin{center}
\resizebox{1.9\columnwidth}{!}{%
  \includegraphics{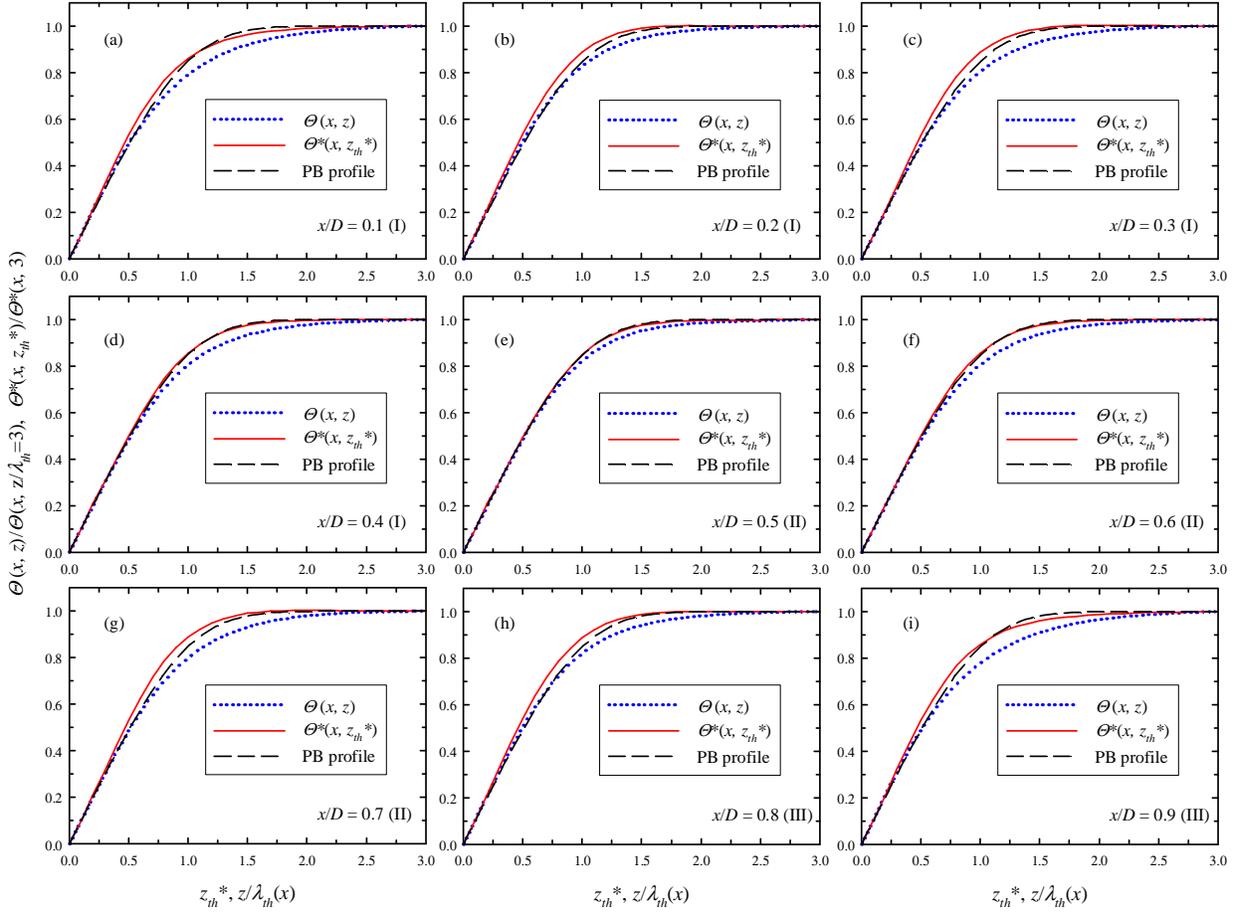}
}
\caption{(color online). Comparison between specifically (see text) normalized vertical temperature profiles obtained at $x/D=0.1$, 0.2, ..., 0.9 near the bottom plate: dynamical $\Theta^*(x,z_{th}^*)/\Theta^*(x,z_{th}^*=3)$ (red solid lines), laboratory $\Theta(x,z)/\Theta(x,z=3\lambda_{th})$ (blue solid lines), and the Prandtl-Blasius laminar thermal profile (black dashed lines), see also figure \ref{fig:fig14}. } \label{fig:fig15}
\end{center}
\end{figure*}

\begin{figure}
\begin{center}
\resizebox{0.95\columnwidth}{!}{%
  \includegraphics{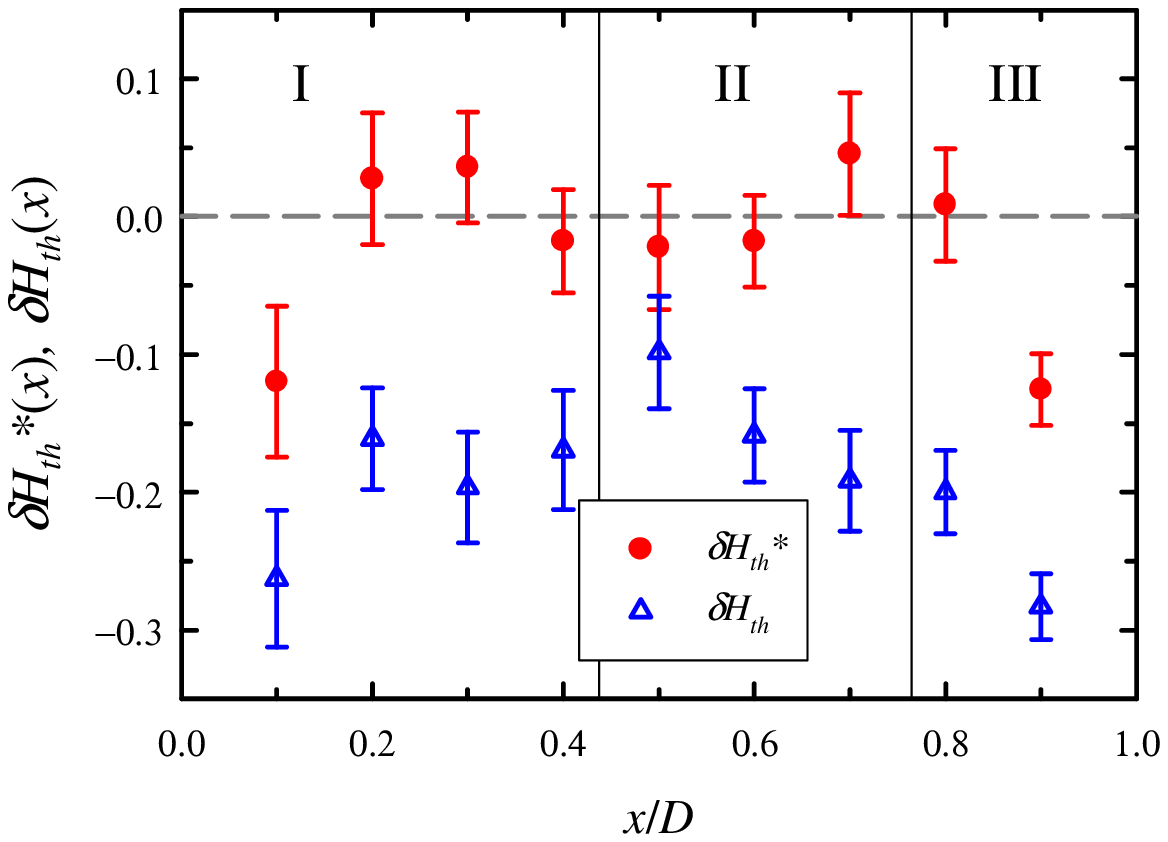}
}
\caption{(color online). The horizontal ($x$-)dependence of the deviations of the local temperature profile shape factors from the Prandtl-Blasius value for the laboratory frame, $\delta H_{th}(x)$ (open triangles) and the time averaged dynamical frame, $\delta H_{th}^*(x)$ (solid circles). Here, both $\delta H_{th}(x)$ and $\delta H_{th}^*(x)$ are calculated based on the profiles in figure \ref{fig:fig15}. The vertical solid lines mark the boundaries at $x_a$ and $x_b$ between the three regions I, II, III.} \label{fig:fig16}
\end{center}
\end{figure}

\section{Temperature profiles near the plate}

We now turn to the corresponding analysis of the temperature field. In Fig. \ref{fig:fig12} we compare the time-averaged thermal BL $z$-profiles obtained in the laboratory frame at nine different horizontal positions with the Prandtl-Blasius profile. The vertical distance $z$ is normalized by the respective local thermal BL thickness $\lambda_{th}(x)$ and the temperature gradient of the Prandtl-Blasius profile is matched to those of the thermal BL profiles. Within region I (Fig. \ref{fig:fig12}(a)), the profiles approach the bulk temperature faster, i.e., they run more quickly to the mean bulk temperature $\Theta=1$ when away from the sidewall. This is because most of the hot plumes rise upwards along the sidewall and hence the influences of thermal plumes on the thermal BL profiles become weaker when away from the sidewall. Figure \ref{fig:fig12}(b) shows the thermal BL profiles obtained at three positions within region II. One can notice that the agreements with the Prandtl-Blasius profile become worse as one moves from the still tilted down flow range of the LSC at the position $x/D=0.5$ to its more plate parallel flow at $x/D=0.7$. This trend is different from what was observed for the kinematic BL profiles shown in Figs. \ref{fig:fig8}(b). -- Finally, the temperature profiles obtained in region III are shown in Fig. \ref{fig:fig12}(c). Both profiles are significantly lower than the Prandtl-Blasius profile for $z/\lambda_{th}(x)\gtrsim0.5$, which we  also attribute to the rising hot plumes in this region.

To calculate the shape factors of the instantaneous local temperature profiles, $H_{th}(x,t)$, we note that the emissions of plumes from the thermal BLs would lead to a much slower approach of the temperature profiles to the asymptotic value, and thus lead to a much lower value of the shape factor, because the temperature adjacent to the BL is not able to immediately relax back to the bulk value when a plume is detaching from the thermal BL. This holds for both the time-averaged as well as the instantaneous profiles, especially for those obtained in the regions near the cell sidewall. For example, see the instantaneous temperature profile obtained at $x/D=0.2$ (red circles) in Fig. \ref{fig:fig6}: here the temperature at the edge of the BL only reaches about $90\%$ of the asymptotic value. Therefore, when calculating $H_{th}(x,t)$ we take the approach of Zhou \emph{et al} \cite{zss2010jfm}: the first maximum temperature near the plate is defined as the asymptotic value of the $z$-profile and is used to normalize the profile.

Figure \ref{fig:fig13} shows PDFs of the instantaneous local thermal shape factors $H_{th}(x,t)$. In the figure, $H_{th}^{PB}=2.61$ is also plotted as the vertical dashed line for comparison. Unlike the case of $H_v(x,t)$ in Fig. \ref{fig:fig9}, the distributions of the $H_{th}(x,t)$ are nearly independent of the horizontal position, i.e., they all collapse on top of each other, except those obtained very near to the two sidewalls ($x/D=0.1$ and 0.9), which are shifted a bit to the left. In addition, the distributions are all peaked close to $H_{th}^{PB}$, indicating that most of the time the instantaneous local temperature profiles over most part of the plate are of Prandtl-Blasius type.

Direct comparisons of the $z$-profiles of the time-averaged local temperature $\Theta(x,z)$ and of the dynamically rescaled field $\Theta^*(x,z_{th}^*)$ with the theoretical Prandtl-Blasius temperature profile at nine different horizontal positions are plotted in Fig. \ref{fig:fig14}. Overall, the mean profiles obtained in the dynamical BL frame, $\Theta^*(x,z_{th}^*)$, are much closer to the Prandtl-Blasius profile than the laboratory frame based profiles, $\Theta(x,z)$, for all horizontal positions $x$, especially for those obtained near the plate's center ($x/D=0.4$, 0.5, and 0.6), which match the Prandtl-Blasius profile exactly.

One noticeable feature of Fig. \ref{fig:fig14} is that the mean temperatures obtained near the cell's sidewall ($x/D\leqslant0.3$ or $x/D\geqslant0.7$) are much lower than the mean bulk temperature $\Theta=1$ even at positions far away from the proper BL range, for both $\Theta(x,z)$ and $\Theta^*(x,z_{th}^*)$. As discussed above, we attribute this to the emissions of thermal plumes, which would lead to a much lower value of the shape factor. Therefore comparison of the shape factors of such profiles with the Prandtl-Blasius value is somewhat meaningless. If, however, we choose the temperature at some position outside of the thermal BL, such as $z/\lambda_{th}(x)=3$ or $z^*_{th}=3$, as the asymptotic value for these positions (rather than the global bulk value) and use the asymptotic temperature to normalize the profiles, the re-scaled mean profiles look quite different. Indeed, following this procedure, we can eliminate the influences of plume emissions on the thermal BL profiles. The obtained mean temperature profiles in the next  figure are much closer to the Prandtl-Blasius type.

Figure \ref{fig:fig15} shows the direct comparison between the various temperature profiles, rescaled in the described way: the dynamical local profiles $\Theta^*(x, z_{th}^*)/\Theta^*(x, z_{th}^*=3)$, the laboratory profiles $\Theta(x, z)/\Theta(x, z=3\lambda_{th})$, and the Prandtl-Blasius laminar thermal BL profile. Again we find a significant preference of the dynamical frame based profiles: Around the thermal BL thickness the laboratory frame based time averaged local profiles $\Theta(x, z)/\Theta(x, z=3\lambda_{th})$ are all much lower than the Prandtl-Blasius profile, while the dynamically rescaled instantaneous local profiles $\Theta^*(x, z_{th}^*)/\Theta^*(x, z_{th}^*=3)$ match the Prandtl-Blasius profile much better. The shape-factor deviations of these rescaled profiles are shown in Fig. \ref{fig:fig16}: The shape-factor deviations $\delta H_{th}(x)$ for the laboratory frame profiles are definitely smaller than zero. In contrast, the dynamical frame based deviations $\delta H_{th}^*(x)$ are much closer to zero, suggesting that our dynamical BL rescaling method can indeed capture the BL properties efficiently.

\section{Conclusions}

In conclusion, we have made a systematic study of the horizontal ($x$-) dependence of the shapes of the $z$-profiles of the kinematic and thermal BLs in turbulent RB convection using 2D numerical data. We have extended our previous studies, which were restricted to the plate's center, to all horizontal positions along the bottom (or top) plate. The major findings can be summarized as follows:

\begin{enumerate}
  \item In situations where the traditional methods based on the time-averaged horizontal velocity profiles are no longer capable of producing physically meaningful BL thicknesses, the time-averaged instantaneous BL thicknesses provide well-defined length scales for both the kinematic and thermal BLs. Such situations can arise, for instance, from the competition between the LSC in the center region and the secondary rolls near the corners.

  \item When the instantaneous local velocity and temperature values are rescaled by their respective instantaneous local BL thicknesses, it is found that the Prandtl-Blasius profiles hold in an instantaneous sense most of the time.

  \item For most parts of the horizontal bottom (or top) plate both the local velocity and temperature profiles match the classical laminar Prandtl-Blasius BL profiles well, if they are re-sampled in the respective dynamically rescaled frames, which fluctuate with the instantaneous local kinematic and thermal BL thicknesses.
\end{enumerate}

\begin{acknowledgments}
We gratefully acknowledge the support of this study by the Natural Science Foundation of China (Nos. 10972229 and 11002085), ``Pu Jiang" project of Shanghai (No. 10PJ1404000), the Shanghai Program for Innovative Research Team in Universities, and E-Institutes of Shanghai Municipal Education Commission (Q.Z.), by the Research Grants Council of Hong Kong SAR (Nos. CUHK403807 and 404409) (K.Q.X.), and by the research programme of FOM, which is financially supported by NWO (R.J.A.M.S. and D.L.).
\end{acknowledgments}


%

\end{document}